\documentclass[aps,prd,final,aps,nofootinbib,showpacs]{revtex4}

\usepackage{graphicx}

\begin{document}


\bibliographystyle{prsty} 

\title{The Periodic Standing-Wave Approximation: Overview
and Three Dimensional Scalar Models}

\author{Zeferino Andrade$^{1}$, 
Christopher Beetle$^{1}$, 
Alexey Blinov$^{1}$, 
Benjamin Bromley$^{1}$, 
Lior M. Burko$^{1}$, 
Maria Cranor$^{1}$, 
Robert Owen$^{1,2}$, 
and Richard H.~Price$^{1}$ }
\affiliation{$^{1}$Department of Physics,
University of Utah, Salt Lake City, Utah 84112, 
$^{2}$Theoretical Astrophysics, California
Institute of Technology, Pasadena, CA 91125}


%

\begin{abstract}
\begin{center}
{\bf Abstract}
\end{center}
The periodic standing-wave method for binary inspiral computes the
exact numerical solution for periodic binary motion with standing
gravitational waves, and uses it as an approximation to slow binary
inspiral with outgoing waves. Important features of this method
presented here are: (i)~the mathematical nature of the ``mixed''
partial differential equations to be solved, (ii)~the meaning of
standing waves in the method, (iii)~computational difficulties, and
(iv)~the ``effective linearity'' that ultimately justifies the
approximation. The method is applied to three dimensional nonlinear
scalar model problems, and the numerical results are used to
demonstrate extraction of the outgoing solution from the standing-wave
solution, and the role of effective linearity.
\end{abstract}

\maketitle
\section{Introduction}\label{sec:intro} 
\subsection*{Background}
The inspiral and merger of a binary pair of compact objects (holes or
neutron stars) is one of the most promising sources of signals
detectable by gravitational wave observatories. For the ground-based
detectors LIGO\cite{gonzalez03}, VIRGO\cite{virgostatus},
GEO600\cite{2000grwa.conf..119L} and TAMA\cite{Ando:1999qj}, binary
merger, especially of intermediate-mass black
holes\cite{Colbert:2002mi} is an exciting possibility; for the
space-based LISA detector\cite{luml.conf,lisawebpage}, the detection
of inspiral/merger of supermassive holes is highly probable, and is
one of the primary scientific targets.

The need for theoretical waveforms for the inspiral/merger has driven
the effort to find a computational solution for the details of the
process, but the difficulty of the task has made this problem also a
measure of the usefulness of numerical computation in general
relativity.  The hope has been that numerical codes evolving initial
data can compute the orbital motion using Einstein's equations and, in
the case of neutron stars, using hydrodynamical equations. These
evolution codes would have, as an intrinsic feature, the loss of
energy by the binary due to outgoing wave energy, and the gradual
inspiral due to this loss.

An important reason for the limited progress on this problem is the
matter of timescales. Near a black hole, the timescale on which the
gravitational field can change is $GM/c^3$, where $M$ is the mass of
the hole; for a neutron star the timescale is several times longer.
The time step in evolution codes is governed by this short timescale.
(More precisely, the spatial grid near the compact objects must be
smaller than $GM/c^2$, and to satisfy the Courant condition, the time
step must be no larger than $1/c$ times this grid size.) By contrast,
the timescale $\sqrt{r^3/GM\; }$, for orbital motion at radius $r$, is
much larger than this, and the timescale for the interesting dynamics,
the radiation-reaction driven inspiral is much greater yet.  The
consequence of this incompatibility of timescales is that a very large
number of time steps is needed in order to see the physics of
interest. And computing a large number of time steps is not yet
possible.  Instabilities\cite{scheel_stable,Yo:2002bm} operating at
the short timescale prevent the code from giving useful answers about
the long timescale.

The origin of the difficulty suggests its cure: an approximation
method that avoids the short timescales. Here we describe such a
method: a solution for periodic sources and fields. We assume that the
compact objects, and their fields, rotate with a constant angular
velocity (to be denoted $\Omega$ below.) This approximation will fail
of course, in the very latest stages of inspiral merger, when the
orbit decays rapidly due to a secular instability or the dynamics of
the final merger.  But that last stage is, by its very nature, rapid;
its timescale is only several times that of the shortest timescale of
the problem. This last stage, then, can plausibly be handled by
numerical evolution codes. Indeed, evolution codes, especially with
perturbation theory handling the final ringdown\cite{laz1}, are already
near to doing this. Our goal, then, is a method that can approximate
the solution up to the time that numerical evolution codes can take
over the task that only they can handle.  Our approach is not
entirely new; it is similar in underlying motivation to a method
introduced by Detweiler and collaborators\cite{det,det2}, but our
approach is very different in its details and its implementation. It
is also very closely related to the approach being used by 
Friedman and his collaborators\cite{friedman}.

Periodic motion and outgoing waves are, of course, impossible in
Einstein's theory, both intuitively and mathematically\footnotemark. For this
\footnotetext{A periodic binary would have no secular change in its
energy, but gravitational waves intuitively remove energy. This
argument can be made mathematically complete using the conservation
law for
 $H^{\alpha\beta\mu\nu}$, as defined in
Misner {\em et al.}\cite{MTW}.}
reason, we will solve for standing waves (to be defined and discussed
below) in the gravitational field. Our periodic standing-wave (PSW)
approach, then, will be to find exact (numerical) periodic
standing-wave solutions of the Einstein field equations and to use these exact
solutions as approximations to the physical problem of slow inspiral
with outgoing waves.

The most basic ideas behind this periodic standing-wave solution have
already been introduced in a previous paper\cite{WKP}, but the
implementation there was applied only to two-dimensional models and
was limited in other ways; in particular, that paper did not discuss the
general meaning of standing waves. An general overview of the PSW project
has also been given\cite{safari}.
Here we present a more specific
discussion, along with numerical results for three-dimensional
models. This paper is meant to serve as the introduction to
the PSW, with subsequent papers presenting more detailed information
on particular methods, and progress on solving the physical problem.

\subsection*{Effective linearity and uses of the method}

A key idea in our approach is the relationship of standing waves to
outgoing waves. In a linear field theory, a definition of standing
waves is that they are half the sum of an outgoing solution and
ingoing solution. Here, as in Ref.~\cite{WKP}, we shall call this sum
LSIO for linear superposition of (half) ingoing and (half) outgoing
solutions.  In a linear theory, such a superposition is itself a
solution.  In our nonlinear field model theories, it will turn out
that --- despite strong nonlinearities --- this continues to be very nearly
true. This {\em effective linearity}, the
approximate equality of the LSIO and a true standing-wave solution, has
already been demonstrated for simple two-dimensional models, and
results for three dimensional nonlinear models will be presented
below. More important, the basis for effective linearity appears
to be robust.  This basis lies in the fact that the strong
nonlinearities in our model theories (and in the physical problem) are
confined to the near-field regions around the sources. In these 
nonlinear near-field regions the solution is insensitive to the
distant boundary conditions; it is substantially the same for ingoing
boundary conditions as for outgoing.  In this near-field region then, the LSIO
will be very nearly a solution despite strong nonlinearities, since we
are superposing nearly identical solutions.  Outside this strong-field
near zone the model theories, and the physical theory, are nearly
linear, so that again the LSIO is a solution. The LSIO will therefore
be a good approximation to a solution everywhere.

Below, we shall choose our definitions of standing-wave solution to be
close to that of a LSIO, and our approximation to a large extent is
based on interpreting a standing-wave solution to be approximately a
LSIO\@.  In the weak field region this LSIO can be deconstructed into
outgoing and ingoing pieces and this deconstruction can be continued
to the strong field source region. (In the source region, the outgoing piece is
simply half the solution.) By doubling the outgoing piece thus
extracted from the standing-wave solution, we thereby arrive at an
approximation to the outgoing solution for a nonlinear problem. It is
in this manner that we will extract an approximation of an outgoing
solution from a computed standing-wave solution.

This is an appropriate place to point out, though not for the last
time, the importance of model problems. In Einstein's theory there is
no obvious meaning to the periodic outgoing solution, so one cannot
make statements about it, let alone carry out numerical studies.
Statements and computations {\em are} possible for nonlinear model
problems, so that tests of effective linearity with such models are
crucial.

The outgoing solutions extracted from our exact periodic solutions can
serve two purposes. First, we can use a quasistationary sequence of
outgoing approximations as a model for the slow physical inspiral.  In
this approach the mass of the system, measured in a weak wave zone far
from the orbiting sources, decreases due to the loss of energy in
outgoing radiation. When we find the system energy as a function of
orbit radius, and we compute the outgoing radiation, we can infer the
rate at which the orbital radius decreases. The difficulty, as with
any such quasistationary sequence, is how to know that we are
comparing the ``same'' system at different radii.  In the case of
neutron star's the answer is clear; baryon number is an unchanging tag
that identifies neutron stars to be the same.  For black holes, the
equivalent tag would be some local mass. The concept of an isolated
horizon\cite{Ashtekar:2000sz} might give us that local mass.

The second use for our extracted outgoing solutions is to provide
initial data for evolution codes. A spacelike slice of our extracted
outgoing field will be an excellent approximation to the physical
initial data, and should be very nearly a solution to the initial
value equations. With little change our extracted outgoing initial
data, can be made into exact (numerical) initial data through the use
of York's decomposition\cite{York71,York73}.

These two purposes of our solution are not distinct. The natural end
point for a quasistationary sequence of PSW solutions is the ``last
orbit,'' the final stage of motion at almost constant radius. This
stage may end due to a secular instability, like that of a particle in
a black hole spacetime, or due to the imminence of the merger, the
formation of the final black hole. In either case, this end point must
be handled by a numerical evolution code, and in either case, the
quasistationary sequence will provide ideal initial data for the
continuation of the problem by numerical evolution.

\subsection*{The nature of the mathematical problem}

In the standard approach to computing binary inspiral, initial data are
evolved forward in time. In our approach, with periodic symmetry
imposed, there is no evolution in the usual sense, and there is not
the usual concept of initial data.  Rather we must satisfy boundary
conditions at large radius: outgoing, ingoing or standing-wave
boundary conditions in model problems, and only standing-wave
conditions in general relativity. The boundary value problems that we
must solve differ in two important ways from common boundary value
problems. First, our partial differential equations (PDEs) are of {\em
mixed} type. They are of elliptical character in some regions and
hyperbolic character in others; this will be particularly clear in the
model problem to be presented below. We will argue that the mixed
character causes no fundamental difficulty, and will demonstrate this
with the model problems. The mixed character, however, does complicate
the use of some of the most efficient numerical means of solving
bounary value problems.  Second, we must define what we mean by
``standing-wave boundary conditions.''  Unlike outgoing and ingoing
conditions, there is no simple local condition corresponding to what
we will mean by standing waves in a nonlinear problem. We will present
two fundamentally different candidates for the standing-wave
condition, and here will present results of computations with one
those of those two choices. (The alternative choice of standing-wave
condition is best implemented with a special numerical method, and
will be presented elsewhere\cite{paperII}.)

Stepping back from such details, one may be led to ask more
fundamental questions about the whole approach. Such  questions arise
especially because the PSW spacetime we compute has some awkward
features. Since the exact PSW solution contains an infinite amount of
gravitational wave radiation, it cannot be expected to be meet the
asymptotic flatness conditions of the theorems about the fall-off of
fields. But the spacetime {\em is} asymptotically flat in that the
spacetime curvature decreases with increasing distance from the binary
source. Another sign that the PSW spacetime has rough edges is that it
must not have regular null infinities; Gibbons and
Stewart\cite{GibbonsStewart} have shown that spacetimes with
well-behaved Scri+ and Scri-- cannot be periodic.

\begin{figure}[ht] 
\includegraphics[width=0.4\textwidth]{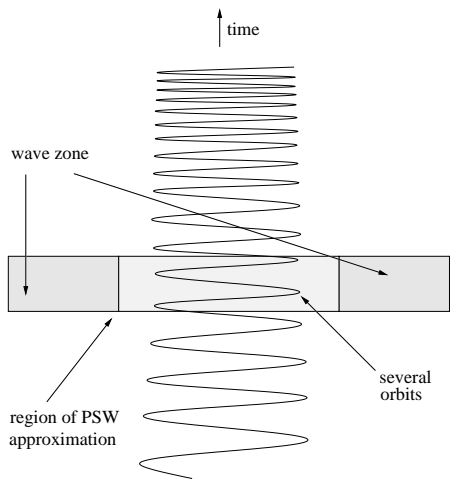}
\caption{
The PSW solution is meant to be an approximation to the 
physical spacetime only in a limited region.\label{fig:PSWpic}}
\end{figure}

It is useful, before diving into details, to clarify what the relationship
is between the slightly singular spacetime we will be computing, and the
physical problem that really interests us. To make this connection we
can think of the binary system going through several orbits at almost
constant radius. A weak wave zone exists at some distance from the
orbit during this epoch of the motion.  The stippled region in
Fig.~\ref{fig:PSWpic} shows the relevant region as part of  the larger physical
spacetime. In this limited region the source motion and the fields are
almost periodic, and it is in this region only that we hope to
approximate the physical fields by the outgoing fields extracted from
the computed PSW solution.  The imperfect asymptotic structure of the
PSW spacetime is therefore irrelevant to its physical usefulness.

In the remainder of this paper we will first present, in
Sec.~\ref{sec:periodic}, the mathematical details of a nonlinear model
with which we clarify many aspects of the PSW approximation. 
We then discuss, in Sec.~\ref{sec:num}, the numerical methods
needed to find PSW solutions, especially those aspects of the 
numerical methods that are idiosyncratic to the special features
(mixed character, standing-wave boundary conditions, nonlinearities)
of our problem. In this section also, results are presented of the 
numerical methods. The results are discussed, and put into the 
context of the next steps in this project\cite{paperII}, in Sec.~\ref{sec:con}.

\section{Periodic solutions, standing waves, and model
problems}\label{sec:periodic}
\subsection*{Mixed PDEs and well-posedness}

\begin{figure}[ht] 
\includegraphics[width=.25\textwidth]{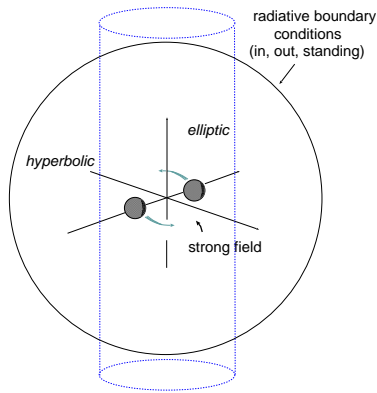}
\caption{The ``light cylinder'' separating the elliptic
and hyperbolic regions of the problem intersects the large
spherical surface on which numerical  boundary conditions 
are imposed.\label{fig:cylandsphere}}
\end{figure}

As stated above, we seek a solution to Einstein's equations in which
the sources and the fields rotate rigidly.  The mathematical statement
of this rigid rotation is that there is a helical Killing vector, a
Killing vector that is timelike close to the sources and spacelike far
from the sources. (For more on helical Killing symmetry see
Ref.~\cite{friedmanuryushibata}.) For fields in flat spacetime our
Killing vector $\vec{\xi}$ takes the form
\begin{equation}
\vec{\xi}=\partial_t+\Omega\partial_\phi
\end{equation}
in spherical or cylindrical spatial coordinates, and 
\begin{equation}
\vec{\xi}=\partial_t+\Omega\left(x\partial_y-y\partial_x
\right)
\end{equation}
in Cartesian spatial coordinates. The parameter $\Omega$, which must
be a constant, can be thought of as the rotation rate of the source
and fields with respect to an inertial reference frame.  For the flat
spacetime case, the null surface $\vec{\xi}\cdot\vec{\xi}=0$ is a
cylinder of radius $1/\Omega$ coaxial with the rotation axis.  (Here
and below we use units in which $G=c=1$.) This cylinder separates the
inner region of timelike $\vec{\xi}$ from the outer region of
spacelike $\vec{\xi}$, as shown in Fig.~\ref{fig:cylandsphere}. Since
this surface, in a sense, represents the points at which the rigidly
rotating fields are moving at $c$, we call this surface the ``light
cylinder,'' in analogy with pulsar electrodynamics.

One immediate advantage of the helical symmetry is that it reduces the number
of independent variables, thereby greatly reducing the computational
difficulty of a problem. In our simple flat spacetime models this
reduction is most easily understood by the fact that helically
symmetric scalars cannot depend in an arbitrary way on the spherical
Minkowski coordinates $t,r,\theta,\phi$ but can depend only on $t$ and
$\phi$ in the combination $\varphi\equiv\phi-\Omega t$. Thus, in the
$t,r,\theta,\varphi$ system the Killing vector is $\vec{\xi}=\partial_t$.
These ideas are clarified with  a simple 
flat spacetime model theory for a scalar field $\Psi$ 
\begin{equation}\label{modelscalar} 
\Psi_{;\alpha;\beta}g^{\alpha\beta}+\lambda F=S\;.
\end{equation}
The term $F(\Psi, x^\alpha)$ is included to
allow for nonlinearity; the constant $\lambda$  adjusts
 the strength of
the nonlinearity.
In order for a helically symmetric solution $\Psi$ to exist, the
explicit coordinate dependence of $F$ must be compatible with the
symmetry. That is, $F$ can have explicit coordinate dependence
only on $r$, $\theta$, and $\varphi$. 
The most natural choice for a
model would be one in which there is no coordinate dependence, one 
in which the background spacetime is featureless. 
We include the  possibility
of spatial dependence for convenience below. Changing the spatial
dependence will help to clarify 
the accuracy of the PSW approximation when nonlinearities are 
important.

In the application of the PSW method to holes, an inner
boundary condition will be used at a small, approximately 
spherical surface. For simplicity here, however, we 
use an explicit source term,
\begin{equation}\label{ptsource} 
S=\frac{\delta(r-a)}{a^2}
\;\delta(\theta-\pi/2)\left[\delta(\varphi)+\delta(\varphi-\pi)
\right]\,,
\end{equation}
representing two points, each of unit scalar charge, in equatorial 
circular orbits, with radius $a$ and angular velocity $\Omega$.
This source term $S$ obeys the symmetry property that is necessary
if  a periodic solution is to exist:
its Lie derivative vanishes along the Killing orbit ${\vec{\xi}}$.

If we are interested only in helically symmetric solutions, then
the field equation (\ref{modelscalar}) reduces to 
\begin{equation}\label{helical3Dscalar} 
\frac{1}{r^2}\frac{\partial}{\partial r}
\left(r^2\frac{\partial\Psi}{\partial r}\right)
+\frac{1}{r^2\sin\theta}\frac{\partial}{\partial\theta}
\left(\sin\theta\frac{\partial\Psi}{\partial\theta}\right)
+\left[\frac{1}{r^2\sin^2\theta}
-\Omega^2
\right]\,\frac{\partial^2\Psi}{\partial\varphi^2}+\lambda
F(\Psi,r,\theta,\varphi)=S(r,\theta,\varphi)\ .
\end{equation}
The mixed character of this PDE shows clearly in the coefficient of
$\partial_\varphi^2\Phi$. The light cylinder is at
$r\sin\theta=1/\Omega$ where this coefficient changes sign.  Inside
the light cylinder ($r\sin\theta<1/\Omega$) the equation is
elliptical; outside it is hyperbolic.  For outgoing solutions of this
equation we impose an outer boundary condition
$\partial_r\Psi=-\partial_t\Psi$, or equivalently
$\partial_r\Psi=\Omega\partial_\varphi\Psi$, on a spherical surface
with a radius large compared to $1/\Omega$. As illustrated in
Fig.~\ref{fig:cylandsphere}, this spherical surface is well outside
the light cylinder in the equatorial plane, so our boundary conditions
are imposed on a surface that passes through both the elliptic and
hyperbolic regions of the problem.

Problems with boundary conditions on closed surfaces are common in the
case of elliptical PDEs.  We argue here that our boundary value
problem with mixed PDEs may be unusual, but is well
posed\cite{ctorre}. Again, a simple model problem will help to clarify
issues. We set the nonlinearity parameter $\lambda$ to zero in
Eq.~(\ref{helical3Dscalar}) so that we can solve the resulting linear
equation as an infinite series. If we choose a Dirichlet condition
\begin{equation}
\left.\Psi\right|_{r=r_{\rm max}}=0\,,
\end{equation}
at a finite radius $r_{\rm max}$, 
then the solution to this linear problem can be written in terms
of spherical Bessel functions $j_\ell,n_\ell$ as
\begin{equation}\label{dirichlet} 
\Psi=\sum_{\ell}\sum_{m={\rm even}}
\frac{2m\Omega Y^*_{\ell m}(\pi/2,0)
Y_{\ell m}(\theta,\varphi)
}{j_\ell(m\Omega r_{\rm max})}
j_\ell(m\Omega r_<)\left[n_\ell(m\Omega r_>)j_\ell(m\Omega r_{\rm max})
-n_\ell(m\Omega r_{\rm max})j_\ell(m\Omega r_>)\right]\ .
\end{equation}
Here $r_<(r_>)$ indicates the smaller (greater) of the quantities $r,a$.
Vanishing of the $j_\ell(m\Omega r_{\rm max})$ denominator means
that the the ``cavity'' $r\leq r_{\rm max}$ has a resonant mode at
frequency $\Omega$. In the case that $r_{\rm max}$ has one of the
resonant values, the solution to the boundary value problem is not
unique. Such values of $r$ are of zero measure, but are dense in the
set of all $r$ choices. This means that the cavity is always
arbitrarily close to a resonance, if sufficiently high angular modes
are computed. A consequence of this is that a numerical computation
does not converge. (Computed solutions depend on the computational
grid size, and become larger with increasing angular resolution.)  The
difficulty is not just one of computational practice. The boundary
value problem is fundamentally ill-posed as a representation of fields
in an infinite space. There is no meaningful $r_{\rm max}\rightarrow\infty$
limit of Eq.~(\ref{dirichlet}).

Problems with mixed elliptic and hyperbolic regions
are of some interest in aerodynamics\cite{tricomi}, but there
are few general results on well-posedness. In those results that
do exist, the nature of the boundary conditions plays a pivotal
role. We have found that this applies to our periodic solutions
also. If we replace the Dirichlet conditions of
Eq.~(\ref{dirichlet}) with the Sommerfeld condition 
\begin{equation}\label{Sommerfeld} 
\left(\partial_r\Psi-\Omega\partial_\varphi
\Psi\right)_{r=r_{\rm max}}=0\,,
\end{equation}
then the problem is found to be well-posed. This is particularly 
clear 
for the linear problem, where the closed form solution 
takes the form 
\begin{equation}
\Psi=\Psi_{\rm out}+\Psi_{\rm extra}\ .
\end{equation}
Here $\Psi_{\rm out}$ is the usual ``outgoing at infinity'' solution
\begin{equation}\label{outinf} 
\Psi_{\rm out}
=\sum_{\ell}\sum_{m={\rm even}}
{-2im\Omega Y^*_{\ell m}(\pi/2,0)
Y_{\ell m}(\theta,\varphi)
}
j_\ell(m\Omega r_<)h^{(1)}
_\ell(m\Omega r_>)
\end{equation}
and 
\begin{equation}\label{extra} 
\Psi_{\rm extra}
\equiv\sum_{\ell}\sum_{m={\rm even}}
{-2im\Omega Y^*_{\ell m}(\pi/2,0)
Y_{\ell m}(\theta,\varphi)
}
\;\gamma_{\ell m}\;
j_\ell(m\Omega r_<)j_\ell(m\Omega r_>)
\end{equation}
with
\begin{equation}
\gamma_{\ell m}\equiv
-\left(\;\frac{h_\ell^{(1)}(z)+i\,dh_\ell^{(1)}(z)/dz}
{j_\ell(z)+i\,dj_\ell(z)/dz
}\right)_{z=m\Omega r_{\rm max}
}\ .
\end{equation}
Since the spherical Hankel function has the asymptotic form
$h^{(1)}_\ell(z)=(-i)^{(\ell+1)}e^{iz} [1/z+{\cal O}(1/z^2 )]$, it follows
that $|\gamma_{\ell m}|$ is of order $1/r_{\rm max}$.  Thus,
$\Psi\rightarrow\Psi_{\rm out}$, as $r_{\rm max}\rightarrow\infty$,
suggesting that the linear problem is well posed\cite{prev}.

Numerical results confirm this suggestion.  With the boundary
condition in Eq.~(\ref{Sommerfeld}), we have encountered no
fundamental difficulty computing convergent solutions to both the
linear and nonlinear versions of Eq.~(\ref{helical3Dscalar}), and have
confirmed that solutions do not depend on the particular (large) value
chosen for $r_{\rm max}$.

\subsection*{Standing waves: iterative method}

The solutions we will be computing in Einstein's theory, of course,
are standing wave solutions, but there are no actual ``standing wave
boundary conditions'' analogous to the Sommerfeld condition in
Eq.~(\ref{Sommerfeld}) for outgoing waves. It is useful, therefore, to
explore the meaning of standing-wave solutions with our model
nonlinear theories.  As pointed out in Sec.~\ref{sec:intro}, our
paradigm for standing waves is the LSIO of a linear theory, the linear
superposition of half ingoing and half outgoing solutions. We shall
extend this definition of standing wave to nonlinear theories in two
ways.  The first is an extension of the Green function method of
Ref.~\cite{WKP}, and is called there the TSGF (time symmetric Green
function) method. For the problem in Eq.~(\ref{modelscalar}) this
method starts by writing the field equation in the form
\begin{equation}\label{iterfirst} 
{\cal L}[\Psi](\Psi)=\sigma_{\rm eff}[\Psi]\,.
\end{equation}
Here the operator ${\cal L}[\Psi]$ depends on $\Psi$ but --- once
$\Psi$ is fixed --- can be considered to be linearly operating on
$\Psi$. Similarly $\sigma_{\rm eff}$ depends on $\Psi$, but --- once
$\Psi$ is fixed--- is to be considered a fixed inhomogeneous term in
the equation, an effective source term. There is no unique way of
putting the field equation into the form of Eq.~(\ref{iterfirst}) for
a nonlinear model problem, or for general relativity. 
The quasilinearity of general relativity, and of our
nonlinear models, means that at least the principal part of ${\cal L}$
is always unambiguous.
There are also
some obvious guidelines to follow. In particular, ${\cal L}$ and
$\sigma_{\rm eff}$ should become $\Psi$-independent in the weak
field limit.

To iterate for an outgoing solution, for example, one would find an
approximate outgoing solution $\Psi^{n}_{\rm out}$, and then would
solve
\begin{equation}\label{itersecond} 
{\cal L}[\Psi^n_{\rm out}](\Psi^{n+1}_{\rm out}
)=\sigma_{\rm eff}[\Psi^{n}_{\rm out}
]\,,
\end{equation}
for outgoing boundary conditions. The result would be the 
improved approximation 
$\Psi^{n+1}_{\rm out}$ to the outgoing waves. 
To find standing waves, this method is modified as follows.
An approximation $\Psi^{n}_{\rm stnd}$ 
is found to the standing-wave 
solution. The equation
\begin{equation}\label{iterlast} 
{\cal L}[\Psi^n_{\rm stnd}](\Psi^{n+1}
)=\sigma_{\rm eff}[\Psi^{n}_{\rm stnd}
]\,,
\end{equation}
is then solved with the outgoing boundary conditions of
Eq.~(\ref{Sommerfeld}) to give $\Psi^{n+1}_{\rm stout}$ and is next solved
with ingoing boundary conditions $\Omega\rightarrow-\Omega$ in
Eq.~(\ref{Sommerfeld})] to give $\Psi^{n+1}_{\rm stin}$. The new
approximation for the standing-wave solution is taken to be
\begin{equation}\label{itstnd} 
\Psi^{n+1}_{\rm stnd}=\textstyle{\frac{1}{2}}\,\Psi^{n+1}_{\rm stout}+
\textstyle{\frac{1}{2}}\,\Psi^{n+1}_{\rm stin}\ .
\end{equation}
We take the $n\rightarrow\infty$ limit of $\Psi^{n+1}_{\rm stnd}$ in
Eq.~(\ref{itstnd}) to be our computed standing-wave solution.

We shall call the iterative method just described ``direct
iteration.''  This sort of direct iteration is useful in solving for
the root of an equation $x=f(x)$ only if $f$ is slowly varying.  In
iteration for $\Psi$ the equivalent condition applies to 
${\cal L}^{-1}\circ\sigma_{\rm eff}$, where ${\cal L}^{-1}$ is the Green
function, the inverse of ${\cal L}$ for the boundary conditions
(ingoing or outgoing) of interest. For direct iteration to converge
the dependence of ${\cal L}^{-1}\circ\sigma_{\rm eff}$ on $\Psi$ must 
be weak and this is the case only for nonlinearities of moderate strength.
For strong nonlinearities another technique must be used.

In Newton-Raphson iteration, one uses the iteration $\Psi^{n}$ to
make linear approximations for ${\cal L}$ and $\sigma_{\rm
eff}$. Equation (\ref{iterfirst})  then takes the form
\begin{equation}\label{NewtRaph} 
{\cal L}[\Psi^n](\Psi)+(\Psi-\Psi^n)\times\left[\frac{\partial{\cal
L}[\Psi]}{\partial\Psi} \right]_{\Psi=\Psi^n}(\Psi^n)=\sigma_{\rm
eff}[\Psi^n] +(\Psi-\Psi^n)\times \left[
\frac{\partial\sigma_{\rm eff}}
{\partial\Psi}
\right]_{\Psi=\Psi^n}\ .
\end{equation}
This equation is linear in $\Psi$, and its solution is taken to be the
next iteration $\Psi^{n+1}$. By choosing appropriate boundary
conditions we can use this scheme to iterate an outgoing or ingoing
solutions. For our standing-wave solution, we follow the same general
scheme as with direct iteration. Using the $n^{\rm th}$ standing-wave
approximation as $\Psi^n$ in Eq.~(\ref{NewtRaph}) we solve using both
in- and outgoing boundary conditions.  As in Eq.~(\ref{itstnd}), the
$n+1^{\rm th}$ standing-wave approximation is taken to be half the sum
of the ingoing and outgoing solutions found this way.

\subsection*{Standing waves: minimization method}

To explain our second,
independent way of defining and computing standing-wave solutions,
it is best to start with the standing-wave solution in the linear
model problem. This is simply half the sum of the ``outgoing
at infinity'' solution in Eq.~(\ref{outinf}), and the equivalent
ingoing solution. The result is 
\begin{equation}\label{swinf} 
\Psi_{\rm stnd}
=\sum_{\ell}\sum_{m={\rm even}}
{2m\Omega Y^*_{\ell m}(\pi/2,0)
Y_{\ell m}(\theta,\varphi)
}
j_\ell(m\Omega r_<)n_\ell(m\Omega r_>)\ .
\end{equation}
In this solution each multipole has an equal amplitude for ingoing and
outgoing amplitudes waves, and one might suspect that this property
suffices to define standing waves for a nonlinear model. This is not,
in fact the case, since we could add a multiple of the homogeneous
solution $j_\ell(m\Omega r)j_\ell(m\Omega a)$ to the $\ell,m$ mode
without changing the balance between ingoing and outgoing.  This
degree of freedom is equivalent to the degree of freedom inherent in
the phase between the outgoing and ingoing waves. This extra degree of
freedom exists also (though not so transparently) in a nonlinear 
problem. 

To resolve this degree of freedom we can use a generalization of a
property that is unique, in the linear case, to the correct
standing-wave solution Eq.~(\ref{swinf}): In each multipole, the
solution is required to have the minimum wave amplitude of any
solution with balanced ingoing and outgoing waves\cite{WBLandP}.

This method, while very interesting in principle, is difficult in
practice to implement in a finite difference boundary value
approach. One could imagine using a guess for the value of a multipole
coefficient at some outer boundary, and then searching for the value
that gives the minimum for the amplitudes of the waves in that
multipole. In a nonlinear problem, the values of each multipole will
influence other multipoles, so the search for minimum waves will, in
principle, be a search in a many dimensional space.

The real difficulty of this numerical approach is that it uses
multipoles as part of the boundary condition. That means that
multipole coefficients must be extracted. Even in
spherical coordinates, the extraction of the multipole coefficients
involves a weighted sum over all angular grid points. Most important,
this sum would not be performed as a postprocessing step on a computed
solution, but rather would have to be written as a set of equations
that would form part of the {\em a priori} problem to be solved. The
set of equations to be solved would then have, in addition to great
complexity, a boundary-related subset connecting distant grid points.
The matrix representation of these equations would not have banded
structure. In addition to these technical difficulties, the use of
spherical coordinates is very ill suited to the structure of our
source objects, so coordinate patches for the sources would be
required.

For these reasons, we have not attempted to use the minimization
criterion in a finite difference code. We have, however, implemented
this criterion with a spectral approach based on a specially adapted
coordinate system. Results from this approach are extremely
encouraging, but the approach poses new computational challenges, so
we are continuing to explore two distinct paths: finite difference 
methods and the iterative definition of standing waves, and 
a spectral/adapted coordinate technique for the minimization 
criterion.  Since the adapted coordinate system necessary for 
the second approach requires a separate development, and is not 
fundamental to the PSW approximation, we confine
the present discussion to the first approach, finite difference 
boundary value problems, with the iterative criterion for 
standing waves.

\section{Numerical implementation and results}\label{sec:num} 

\subsection*{Extraction of an outgoing approximation}
Model problems allow us to test a key idea of the PSW approach, that a
good approximation of the outgoing solution can be extracted from the
computed standing-wave solution
\begin{equation}\label{extra1} 
\Psi_{\rm stndcomp}
=
\sum_{\mbox{even $\ell$}}
\ \
\sum_{m=0,\pm2,\pm4..}
\alpha_{\ell m}(r)Y_{\ell m}(\theta,\varphi)\,.
\end{equation}
The coefficients $\alpha_{\ell m}(r)$ are computed from $\Psi_{\rm
comp}$, by projection with $Y^*_{\ell m}$. From the reality of
$\Psi_{\rm stndcomp}$, the coefficients will obey $\alpha_{\ell
m}^*=\alpha_{\ell -m}$, and from the standing-wave symmetry
($\cos{m\varphi}$ only, no $\sin{m\varphi}$ terms) they will also obey
$\alpha_{\ell m}=\alpha_{\ell -m}$.

This form of the computed standing-wave solution is compared with a
general homogeneous linear ($\lambda=0$) standing-wave (equal
magnitude in- and outgoing waves) solution of
Eq.~(\ref{helical3Dscalar}), with the symmetry of two equal and
opposite sources:
\begin{equation}
\Psi_{\rm stndlin}=
\sum_{\mbox{even $\ell$}}
\ \
\sum_{m=0,\pm2,\pm4..}
Y_{\ell m}(\theta,\varphi)\,\left[
\textstyle{\frac{1}{2}}\;
C_{\ell m}h^{(1)}
_{\ell}(m\Omega r)
+\textstyle{\frac{1}{2}}\;C^*
_{\ell m}h^{(2)}
_{\ell}(m\Omega r)
\right]\,,
\end{equation}
where $C_{\ell-m}=C_{\ell m}^*$, from the reality of $\Psi_{\rm
stndlin}$. A fitting, in the weak-field zone, of this form of the
standing-wave multipole to the computed function $\alpha_{\ell m}(r)$
gives the value of $C_{\ell m}$.

By viewing the linear solution as half-ingoing and half-outgoing
we define the extracted outgoing solution to be
\begin{equation}\label{extractfirst} 
\Psi_{\rm exout}=
\sum_{\mbox{even $\ell$}}
\ \
\sum_{m=0,\pm2,\pm4..}
Y_{\ell m}(\theta,\varphi)\,
C_{\ell m}h^{(1)}
_{\ell}(m\Omega r)\ .
\end{equation}
Since this extracted solution was fitted to the computed solution
assuming only that linearity applied, it will be a good approximation
except in the strong-field region. In the problems of interest, the
strong-fields should be confined to a region near the sources. In
those regions, small compared to a wavelength, the field will
essentially be that of a static source, and will be insensitive to the
distant radiative boundary conditions. As pointed out in
Sec.~\ref{sec:intro}, the solutions in this region will be essentially
the same for the ingoing, outgoing, and standing-wave problem.
In this inner region then, we take our extracted outgoing solution
simply to be the computed standing-wave solution, so that
\begin{equation}
\Psi_{\rm exout}=\left\{
\begin{array}{ll}
\sum Y_{\ell m} C_{\ell m}h^{(1)}_{\ell}\ \ &\mbox{weak field outer region}
\\
\Psi_{\rm stndcomp}
&\mbox{strong field inner region}
\end{array}
\right.\ .
\end{equation}

The boundary between a strong field inner region and weak field outer
region would ideally be a closed surface surrounding each of the
source regions. This is easily implemented with the adapted
coordinates to be introduced in a subsequent paper. Here, for
simplicity, we take the boundary to be a spherical surface around the
origin. In order for the extracted solution to be smooth at this
boundary, we use a blending of the inner and outer solution in a
transition region extending between radii $r_{\rm low}$ and 
$r_{\rm high}$ and, in this region, we take 
\begin{equation}\label{extracteq} 
\Psi_{\rm exout}=\beta(r)\;\sum Y_{\ell m} C_{\ell m}h^{(1)}_{\ell}
+[1-\beta(r)]\;\Psi_{\rm stndcomp}\,.
\end{equation}
Here
\begin{equation}\label{extractlast} 
\beta(r)\equiv  
3\left[\frac{r-r_{\rm low}}{r_{\rm high}-r_{\rm low}}\right]^2
-2\left[\frac{r-r_{\rm low}}{r_{\rm high}-r_{\rm low}}\right]^3
\,,
\end{equation}
so that $\beta(r)$ goes from 0 at $r=r_{\rm low}$ to unity at
$r=r_{\rm high}$ and has a vanishing $r$-derivative at both ends.

In the case of our typical choice $\Omega=0.3$, the
value
 of $r_{\rm low}$ is chosen to be $r=1.3\,a$ the value at which 
the static and standing-wave solutions 
of the linear problem differ by 10\%. This value should decrease
with increasing $\Omega$, but it must be larger than the orbital
radius $r=a$, so we choose it to be
\begin{equation}
r_{\rm low}
=a[1+0.3(0.3/\Omega)]\,.
\end{equation}
In order to have a moderately thin transition region
we somewhat arbitrarily take
\begin{equation}\label{extralast} 
r_{\rm high}
=a[1+0.6(0.3/\Omega)]\,.
\end{equation}
For the numerical results reported below, the extraction details
of Eqs.~(\ref{extra1})--(\ref{extralast}) are used, and extraction
is carried out using the $\ell=0,2,4$ multipoles.

\subsection*{Choice of model}
To verify and demonstrate several innovative features (well-posed
mixed boundary value problem, standing waves, effective linearity) of
the PSW approximation, we use the nonlinear scalar model of
Eq.~(\ref{helical3Dscalar}), with the $\delta$ function sources given by
Eq.~(\ref{ptsource}). We make the simplifying assumption that 
the nonlinear function $F$ in Eq.~(\ref{helical3Dscalar})
depends only on $\Psi$, not on its derivatives. From this
an obvious simplification follows for 
the  iteration method of 
Eqs.~(\ref{iterfirst}) --
(\ref{itstnd}).
We replace Eq.~(\ref{iterfirst}) by
\begin{equation}\label{ourprob} 
{\cal L}(\Psi)=\sigma_{\rm eff}[\Psi]\,.
\end{equation}
with 
${\cal L}$ taken to be
\begin{equation}\label{ourL} 
{\cal L}\equiv 
\frac{1}{r^2}\frac{\partial}{\partial r}
\left(r^2\frac{\partial}{\partial r}\right)
+\frac{1}{r^2\sin\theta}\frac{\partial}{\partial\theta}
\left(\sin\theta\frac{\partial}{\partial\theta}\right)
+\left[\frac{1}{r^2\sin^2\theta}
-\Omega^2
\right]\,\frac{\partial^2}{\partial\varphi^2}\,.
\end{equation}
The effective source
term includes both   the true point source 
and the nonlinear term
\begin{equation}\label{oursigma} 
\sigma_{\rm eff}[\Psi]=
\;\frac{\delta(r-a)}{a^2}
\delta(\theta-\pi/2)\left[\delta(\varphi)+\delta(\varphi-\pi)
\right]-\lambda F\,.
\end{equation}

Our choice of the nonlinearity function $F$ is
\begin{equation}\label{modelF} 
F=
\frac{\Psi^5}{\Psi_0^4+\Psi^4}\,.
\end{equation}
(We will comment below on the difference between this choice and that
made in previous work, including previous versions of this paper.)
Here $\Psi_0$ is a second nonlinearity parameter ($\lambda$ being the
first). We shall choose $\Psi_0$ to be less than unity; in the
numerical results to be presented, $\Psi_0$ is taken to be
$0.15$.

To understand the effect of this nonlinearity, let $R$ denote the
distance from one of the point sources. Very near a  source point, at
very small $R$, where the field is strong, $F$ has the limit
$F\rightarrow\Psi$, so that the solution of
Eqs.~(\ref{ourprob})--(\ref{modelF}) approximately has  the Yukawa form
\begin{equation}\label{yukawa} 
\Psi\approx \frac{e^{-\sqrt{-\lambda\;}R}
}{4\pi R}\quad\quad\quad\mbox{\rm near source pt}\,.
\end{equation}
At some distance from the source ---
call it $R_{\rm lin}$ --- the
field $\Psi$ becomes smaller than $\Psi_0$, and $F$ can be
approximated as $\Psi^5$. Since $\Psi$ itself is less than $\Psi_0$,
and hence less than unity, this $\Psi^5$ nonlinearity is small enough
to be considered a perturbative correction.

If the transition at $R_{\rm lin}$ takes place well inside the near
zone of the problem, then the effect of the nonlinearity can be
understood as follows: Near a source point the solution has the form
of a unit strength Yukawa potential. At distance $R_{\rm lin}$ the
effect of the $\lambda F$ term is turned off and the solution becomes
a simple Coulomb potential. The source strength for this Coulomb
field, though, will be less than unity. Due to the
$\exp{(-\sqrt{-\lambda}R)}$ Yukawa factor, the source strength
decreases in the region from $R=0$ to $R=R_{\rm lin}$, and the effect of the
nonlinearity is to reduce the effective source strength by a factor of
order $\exp{(-\sqrt{-\lambda}R_{\rm lin} )}$. Since this transition
takes place well within the near zone, it should be this {\em reduced}
source strength that is responsible for generating radiation. The
effect of the nonlinearity on radiation, then, will be the same
reduction factor $\exp{(-\sqrt{-\lambda}R_{\rm lin} )}$, and we can
easily estimate the size of this nonlinear effect. One estimate
can be found by solving 
\begin{equation}\label{estimate1} 
\Psi\approx \frac{e^{-\sqrt{-\lambda\;}R_{\rm lin}
}}{4\pi R_{\rm lin}}=\Psi_0
\end{equation}
for $R_{\rm lin}$, and using this value of $R_{\rm lin}$ in the
expression $\exp{(-\sqrt{-\lambda}R_{\rm lin} )}$ for the reduction
factor. Another estimate follows by solving the spherically
symmetric static nonlinear problem for a unit strength source
\begin{equation}\label{estimate2} 
\frac{1}{r^2}\frac{\partial}{\partial r}
\left(r^2\frac{\partial \Psi}{\partial r}\right)
+\lambda\;\frac{\Psi^5}{\Psi_0^4+\Psi^4}=\delta^3({\vec{r}})\,.
\end{equation}
(Here the right hand side is the unit $\delta$ function at the origin.)
For this solution the ratio is found of the large-$r$ monopole moment
to the small-$r$ monopole moment, and this ratio is taken as the
reduction factor. Since these methods for the reduction factor ignore
the nonlinear interaction between the two source points, and since they
assume that all the wave generation occurs far outside  $R_{\rm lin}$,
they can only be considered an approximation for the nonlinear reduction
effect on the wave amplitude. We shall see, however, that these estimates
are
accurate
enough to be taken as a good heuristic explanation of the role of the
nonlinearity.

In previous work, a form of the nonlinearity was used that was
different from that in Eq.~(\ref{modelF}).  To give that previous form
we first defined the distance $R_+$ ($R_-$) from the source point on
the $x$ axis at $x=a$ ($x=-a$) to be given by
\begin{equation}
R_\pm^2=(r\sin\theta\cos\varphi\mp a)^2+r^2
\sin^2\!\theta\cos^2\!\varphi+r^2\cos^2\!\theta\,.
\end{equation}
We then introduced the distance variable
\begin{equation}
\chi\equiv\sqrt{R_+R_-\;}\,.
\end{equation}
At either of the source points $\chi\rightarrow0$, and far from the
sources $\chi\rightarrow r$. In terms of $\chi$, the form of 
the nonlinearity previously used\cite{earlierF} is 
\begin{equation}\label{prevmodelF} 
F_{\rm prev}
=\left(\frac{\chi}{na}\right)^n e^{(n-\chi/a)}\;
\frac{\Psi^3}{1+\Psi^2}\,.
\end{equation}
The $\chi$-dependent prefactor $(\chi/na)^ne^{(n-\chi/a)}$ was
included so that we could force the nonlinearity to be concentrated
near $\chi=na$. By choosing $n$ to be 5 or 10 we could, in this way,
have strong nonlinearity in the wave zone, and we could numerically
demonstrate the failure of effective nonlinearity. 
The $\chi$-dependent prefactor, however, makes it 
difficult to find a numerical solution that is physically
meaningful.

The prefactor is a difficulty because the solution near the source can
have either the Yukawa form $\exp{(-\sqrt{-\lambda}R)}$, or the
``anti-Yukawa'' form $\exp{(+\sqrt{-\lambda}R)}$. If there is any of
the latter included in the solution, then the field gets larger at
larger distances from the sources, so the strong nonlinearity is never
suppressed, the $\lambda F$ term continues to approximate
$\lambda\Psi$, and the sum of the Yukawa and anti-Yukawa forms
continues to be a valid solution. But if the anti-Yukawa part is
present, the solution cannot meet the fall-off conditions  at an outer
boundary at large $r$. Without the prefactor, then, the outer boundary
conditions act to suppress the anti-Yukawa part of a solution.  With
the prefactor present, however, the nonlinearity can be turned off by
the fall off of $\exp{(-\chi/a)}$, even if the solution contains an
anti-Yukawa part close to the sources.  The prefactor, in effect,
shields the inner region from the influence of the  outer
boundary conditions. When the prefactor is included in the
nonlinearity, the solution in the inner region will be a somewhat
unpredictable mixture of Yukawa and anti-Yukawa parts that is
sensitive to grid spacing.

The choice made for the $\Psi$ dependence in Eq.~(\ref{modelF}), rather
than that in Eq.~(\ref{prevmodelF}), is motivated by the fact that
$F\sim\Psi^3$ falls off rather slowly in the weak wave zone. Changing
the form of $F$ to $\Psi^5/(1+\Psi^4)$ cures this slow fall off, but
imposes a very sharp cutoff near the sources, one that is too sharp
for our relatively coarse computational grid. By taking $F$
proportional to $\Psi^5/(\Psi_0^4+\Psi^4)$, with a fairly small value
of $\Psi_0$, the falloff of $F$ is smoothed
out and moved to a larger distances from the source.

\subsection*{Numerical methods}
Since ${\cal L}$ is independent of $\Psi$ we can (as in
Ref.~\cite{WKP}) compute once and for all the inverses of ${\cal L}$,
i.e.\,, the Green functions corresponding to specific boundary
conditions. In this way, we can compute ${\cal L}^{-1}_{\rm out}$
and ${\cal L}^{-1}_{\rm in}$, the Green functions
for outgoing and for ingoing boundary conditions. The 
direct iterative method of Eqs.~(\ref{itersecond}) -- (\ref{iterlast}), 
then amounts to
\begin{equation}\label{it4outin} 
\Psi^{n+1}_{\rm out}
={\cal L}^{-1}_{\rm out}\left(\sigma_{\rm eff}
[
\Psi^{n}_{\rm out}]\right)
\quad\quad\quad\quad
\Psi^{n+1}_{\rm in}
={\cal L}^{-1}_{\rm in}\left(\sigma_{\rm eff}[
\Psi^{n}_{\rm in}]\right)
\end{equation}
 \begin{equation}\label{it4stnd} 
\Psi^{n+1}_{\rm stnd}
=\textstyle{\frac{1}{2}}\left\{
{\cal L}^{-1}_{\rm out}+{\cal L}^{-1}_{\rm in}
\right\}\sigma_{\rm eff}[
\Psi^{n}_{\rm stnd}]\,.
\end{equation}
Since ${\cal L}$ has no $\Psi$ dependence, the basic Newton-Raphson
iteration simplifies to 
\begin{equation}\label{simpNR} 
\left\{{\cal L} -\left[
\frac{\partial\sigma_{\rm eff}}
{\partial\Psi}
\right]_{\Psi=\Psi^n}\right\}  \Psi
=\sigma_{\rm
eff}[\Psi^n] 
-\Psi^n\;\left[
\frac{\partial\sigma_{\rm eff}}
{\partial\Psi}
\right]_{\Psi=\Psi^n}\ .
\end{equation}
This Newton-Raphson approach can be applied to find outgoing, ingoing and
standing-wave solutions analogous to Eqs.~(\ref{it4outin}) and (\ref{it4stnd}).

Each iteration of Eqs.~(\ref{it4outin}), (\ref{it4stnd}) or
(\ref{simpNR}) is equivalent to the solution of a large set of linear
equations.  Such systems are most typically encountered for elliptic
boundary value problems, and are typically solved most efficiently
with relaxation methods, or related methods (e.g.\,, multigrid) based
on the geometry of the problem. Such methods start with an approximate
set of values for each of the unknowns at every point of the numerical
grid. At each point the solution is then recalculated on the basis of
the values at nearby grid points. This method sweeps through all the
points of the grid and is iterated until an error criterion is
met. Such a method must be compatible with the domain of dependence
for the points of the grid. For an elliptic PDE, for example, the
values of unknowns are updated at a central point of a set of grid
points.  For a hyperbolic PDE, on the other hand, the field
computation, or updating, must be done only at a point in the
``future'' of those grid points being used. For a mixed boundary value
problem a relaxation method has special difficulties, especially at
the interface between elliptic and hyperbolic regions. Nevertheless,
relaxation methods have been successfully applied to mixed PDEs in
transonic aerodynamics, first by Murman and Cole\cite{MurmanCole}. The
slow convergence of this method at the interface (the ``sonic
surface'' in transonic aerodynamics) can be improved with special
techniques that may need to be specific to the problem\cite{Nixon}.

We are presently investigating relaxation and other numerical methods
(e.g.\,, decomposing the grid into regions and applying different
techniques, preconditioners, etc.) for large grids and many variables.
For our three-dimensional scalar problem illustrated here, however, we
have been able to use a more-or-less straightforward method of
inverting the matrix for the finite difference equations.  

In one approach to finding an iterative solution, we use matrix
inversion at each step of the direct iteration of
Eqs.~(\ref{it4outin}), (\ref{it4stnd}), and we take advantage of the
fact that ${\cal L}^{-1}$ is rotationally symmetric (i.e.\,, it is
translationally symmetric with respect to $\varphi$), and we work with
the Fourier components $\Psi^{n}_m(r,\theta)e^{im\varphi}$ of the
iterative solution.  At each step of iteration we project out the
Fourier components of the effective source. Due to the nonlinearity in
the effective source, the Fourier modes of
$\Psi^{n}_m(r,\theta)e^{im\varphi}$ mix in this step, but ${\cal
L}^{-1}$ is rotationally symmetric so the Fourier modes do not mix in
the step of solving for $\Psi^{n+1}$.  This method takes advantage of
the efficiency of a fast Fourier transform (FFT) and reduces the RAM
needed to little more than that for a two-dimensional $r,\theta$ grid.
This method, therefore, allows a rather fine grid in $r$ and
$\theta$. 

We have used this efficient FFT method extensively, but direct
iteration has the drawback already cited following Eq.(\ref{itstnd}):
it is limited in the strength of the nonlinearity it can
handle. Direct iteration will not converge for very strong
nonlinearity. The iterative Newton-Raphson method of
Eq.~(\ref{simpNR}), on the other hand, does almost always converge
once one has a solution sufficiently close to the correct
solution. The operator on the left in Eq.~(\ref{simpNR}), however,
contains the previous iteration $\Psi^n$, which is not symmetric in
$\varphi$, so that the FFT method cannot be used with Newton-Raphson
iteration. This has meant that a relatively coarse grid, or large RAM,
had to be used. We have not yet implemented a parallelizable method
for solving the iteration steps, and have been restricting most runs
to 8GB\@.

It is worth mentioning an interesting hybrid method that we have
explored. The problem in Eqs.~(\ref{ourprob})--(\ref{modelF}),
outside the point sources, can be written as
\begin{equation}\label{lambdaleft} 
\left({\cal L}+\lambda\right)\Psi=\lambda
\Psi_0^5\;\frac{\Psi}{\Psi_0^4+\Psi^4}\,.
\end{equation}
The nonlinearity on the right is never large; it is small both for
$\Psi>\Psi_0$ near the sources, and for $\Psi<\Psi_0$ far from the
sources. The weakness of the formal nonlinearity suggests that a
solution may converge with direct iteration even for large nonlinear
effects. The operator $\left({\cal L}+\lambda\right)$, furthermore, is
rotationally symmetric, so the FFT method can be used. 
The method, however, turns out to have a serious flaw.  Where
$\Psi$ is small, the left-hand side of Eq.~(\ref{lambdaleft}) is
dominated by $\lambda\Psi$, which is very nearly equal to the
right-hand side. In the analogy we gave, following Eq.(\ref{itstnd}), to
the iterative solution for a root of $x=f(x)$, this is equivalent to
$f$ having a derivative very close to unity at the root.
It would appear that this difficulty could be avoided by iterating
Eq.~(\ref{lambdaleft}) near the source, where the nonlinearity is
strong, and iterating the standard form of the problem in the weak
field region. Numerical experiments with this approach have been
inconclusive. Since we do not intend to use a single-patch spherical
coordinate sysem in the future, we have not examined this hybrid
method exhaustively.

In practice we have used the following eclectic approach to find
solutions: (i)~For linear models, for which no iteration is required,
we have taken advantage of the RAM-reduction of the FFT
method. (ii)~For strongly nonlinear models we have used Newton-Raphson
iteration on a three-dimensional (non-FFT) grid, and have used
continuation (i.e.\,, ``ramping up'') both in $\lambda$ and in
$\Omega$. Despite RAM limitations, we have been able to confirm that
the solutions are second-order convergent.  (iii)~For $-\lambda$ less
than around 2, it has been been possible to find solutions with the
direct-iteration, FFT method. These solutions have been compared with
the corresponding solutions from the non-FFT, Newton-Raphson method,
and have been found to agree within the numerical uncertainty in the
solutions.

In applying the iteration methods, and looking for convergence, we
have used two error measures. One, $\epsilon_{\rm iter}\equiv$
$\Psi^{n} -\Psi^{n-1} $, is the difference at a grid point between the
computed value at a grid point, and the value computed at the previous
iteration. The second error measure, $\epsilon_{\rm soln}$ is the
value of $\left({\cal L}(\Psi^n)-\sigma_{\rm eff}[\Psi^n]\right)$ at a
grid point.  In our FFT computations the criterion for convergence was
to have the rms value of $\epsilon_{\rm iter}$ (averaged over the
entire grid) fall below $1\times10^{-6}$. The value of $\epsilon_{\rm
soln}$ was also monitored in the FFT computations and was found not to
be larger than $1\times10^{-6}$ at any grid point, and to have an rms
value typically around $1\times10^{-7}$. A much more stringent
requirement for convergence was used in the Newton-Raphson
computations: the rms value of $\epsilon_{\rm soln}$ had to fall below
$5\times10^{-11}$ for the solution to be acceptable.

\begin{figure}[ht]
\begin{center}
\includegraphics[width=0.8\textwidth]{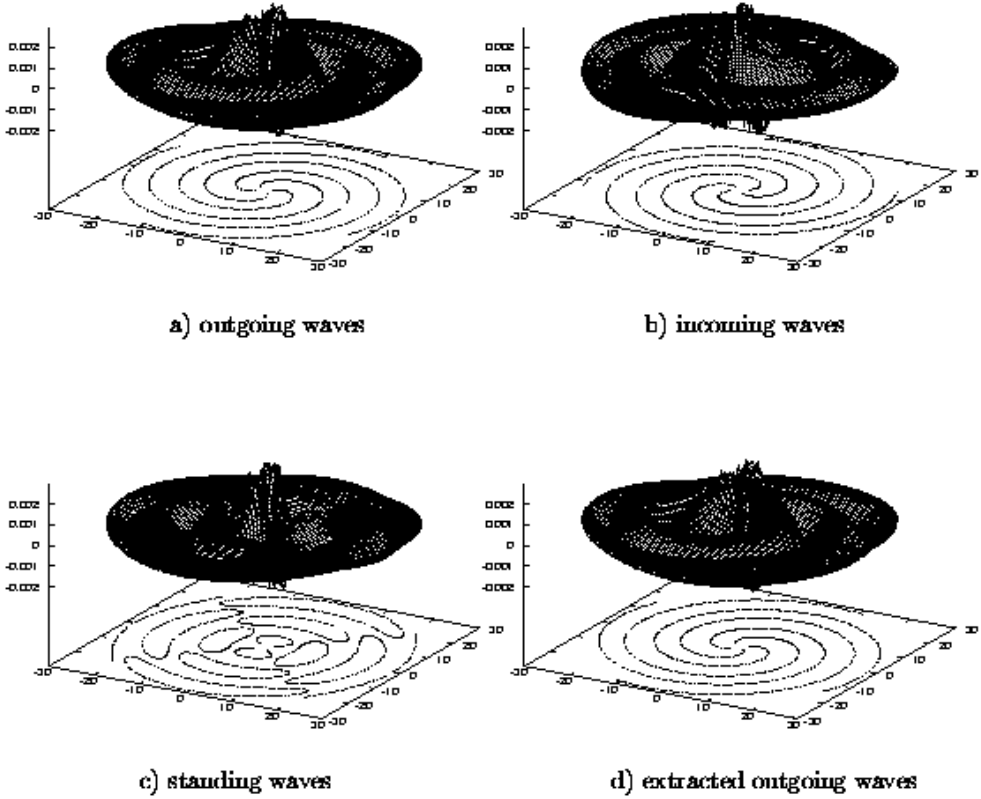}
\end{center}

\vspace{-1.6in}
\caption{\label{fig:eqplane} The $\Psi$ field for two rotating point
sources in the equatorial plane.  The fields shown are nonlinear
solutions of Eqs.~(\ref{ptsource}), (\ref{helical3Dscalar}) and
(\ref{modelF}), with $a\Omega=0.3$ and $\lambda=-1$.  For
clarity, the $\varphi$-average is removed at each radius.  Parts (a)
and (b) of the figure show, respectively, the nonlinear outgoing and
ingoing solutions. Part (c) is the standing-wave solution, and
part (d) is the outgoing solution extracted from it. The
vertical scale gives field strength (arbitrary units) and the
horizontal coordinates are corotating Cartesian coordinates in units
of $a$, the distance of a source from the rotation axis.  }
\end{figure}

\subsection*{Numerical results}
We first illustrate the fundamental concept of the PSW method with
various solutions of Eq.~(\ref{helical3Dscalar}), with the
nonlinearity given in Eq.~(\ref{modelF})\,.  Figure \ref{fig:eqplane}
shows solutions in the equatorial ($\theta=\pi/2$) plane; the
amplitude of the field $\Psi$ is plotted as a function of corotating
Cartesian coordinates $x=r\cos\varphi$, and $y=r\sin\varphi$. The
source points are on the $x$ axis at $x=\pm a$, and the outer
boundary  is at $r=30a$.  For all four plots,
$a\Omega=0.3$, $\lambda=-1$ and $\Psi_0=0.15$.  The results plotted
are those from direct iteration with the FFT method, for a
computational grid using 361 radial divisions, 16 divisions in
$\theta$, and 32 Fourier modes. For all models, the computed results
are dominated by the monopole, so for clarity in the figures the
$\varphi$-average of the solution has been subtracted at every radius.
It is worth noting that this procedure not only removes the monopole
(the $\ell=0$ part of the solution), but also removes the $m=0$ part
of the quadrupole, etc.

The plot in part (a) of Fig.~\ref{fig:eqplane} shows the outgoing
solution (solution for outgoing boundary conditions); the plot in part
(b) shows the corresponding ingoing solution. The plot in part (c) is
the computed standing-wave solution for the same problem parameters.
(Note: this nonlinear standing-wave solution is {\em not} half the
superposition of the outgoing and ingoing solutions.  Rather, it is
the nonlinear field equation solved with the standing-wave definition
discussed in Sec.~\ref{sec:periodic}.)  Part (d) shows the key idea of
the PSW approximation, the outgoing solution extracted from the
standing-wave solution, by the extraction method described in
Eqs.~(\ref{extra1})--(\ref{extralast}).  When the PSW method is used
in general relativity, it will be possible only to compute the
standing-wave solution; the extracted outgoing solution will represent
the approximation to the physical, outgoing solution.

Table~\ref{table1} gives quantitative results for strongly nonlinear
outgoing waves. In that table, values are given for the reduction
factor due to the nonlinearity. As explained following
Eq.~(\ref{yukawa}), this is the factor by which the 
nonlinearity decreases the
amplitude of the
waves. (For the same $a\Omega$ and
source strength, the amplitude of outgoing waves for the linear
problem $\lambda=0$ problem is compared to the  amplitude for a
problem with $\lambda\neq0$.) The fact that the reduction factors are
significantly different from 100\% shows that we are able to compute
models in which nonlinear effects are strong.  In the table the
computed reduction is compared with estimates from heuristic models of
Eqs.~(\ref{yukawa})--(\ref{estimate2}) in which $\Psi$ is taken to
have a Yukawa form very near the source, and a Coulomb form further
out, but well within the near zone. The agreement of the computations
with the estimates is strong evidence that the heuristic model
captures much of the nature of the nonlinear effect.
\begin{table}
\caption{The reduction factor for outgoing waves due to the nonlinearity. 
For all cases, $\Psi_0=0.15$.
The second column refers to Eq.~(\ref{estimate1}). This equation
is solved for $R_{\rm lin}$. Estimate 1 uses this value of $R_{\rm lin}$
in
$\exp{(-\sqrt{-\lambda}R_{\rm lin} )}$. Estimate 2 is the 
reduction factor found from a numerical solution of 
Eq.~(\ref{estimate2}). The last column gives the results
from Newton-Raphson computation with $a\Omega=0.3$, with the outer boundary
at $r=30a$, and with a  $r,\theta,\varphi$ grid of
$120\times20\times32$.
\label{table1} }
\begin{tabular}{|c|c|c|c|}
\hline
$\lambda$&Estimate 1&Estimate 2&Computation\\
\hline
-1&69\%&87\%    &78\%\\
-2&62\%&73\%    &68\%\\
-5&53\%&65\%    &55\%\\
-10&46\%&54\%   &47\%\\
-25&37\%&41\%   &35\%\\
\hline
\end{tabular}
\end{table}

Table \ref{table2} gives information on the numerical errors of the
most computationally intensive solution type: that for strongly
nonlinear waves computed via Newton-Raphson iteration. A single
physical model ($\lambda=-10$, $\Psi_0=0.15$, $a\Omega=0.3$, outer
boundary at $r=30a$) is computed on five different grids.  As a
measure of the truncation error for grid $k$, the L2 difference (the
square root of the average square difference) is found between the
results for grid $k$ and for grid $k+1$ . This is listed in Table
\ref{table2} as the error in grid $k$.
These results, especially for the finest three grids, suggest
quadratic convergence of the numerical process.
\begin{table}
\caption{Convergence of finite difference computations.  Nonlinear
outgoing solutions are computed with five different grid resolutions
for $\lambda=-10$, $\Psi_0=0.15$, $a\Omega=0.3$ and outer boundary at
$r=30a$. An $L2$ norm is computed for the difference between the
solution for grid $k$ and grid $k+1$. This is reported as the
``Error'' for grid $k$.  \label{table2}}
\begin{tabular}{|c|c|c|}
\hline
$k$&$n_{r}\times n_{r}\times n_{\varphi}$&Error\\
\hline
1&  90$\times$10$\times$16  & 2.71\,E-5\\
2& 120$\times$14$\times$22  & 1.60\,E--5\\
3& 150$\times$16$\times$26  & 8.68\,E-6\\
4& 180$\times$20$\times$32  & 5.22\,E-6\\
5& 210$\times$24$\times$38   &\\
\hline
\end{tabular}
\end{table}

The crux of the PSW method is that a good approximation to a nonlinear
outgoing solution can be extracted from a standing wave nonlinear
solution.  Examples of this are given in the next two figures, the
central numerical results in this paper.

\begin{figure}[ht]
\begin{center}
\includegraphics[width=0.4\textwidth]{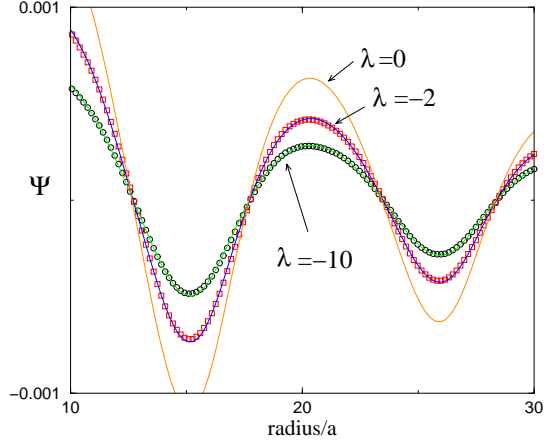}
\end{center}
\caption{\label{fig:comp0line} Extracted outgoing nonlinear waves
vs.~true outgoing nonlinear waves.  For $\lambda=0,-2,-10$,
$\Psi_0=0.15$, $a\Omega=0.3$, with a $180\times20\times32$ grid.  The
field $\Psi$ is shown as a function of r along a radial line through
the source point, i.e.\,, along the $\theta=\pi/2$, $\varphi=0$ line.
Continuous curves show computational results for outgoing waves.
Discrete points, for the nonlinear models, show the approximate
outgoing waves extracted from standing wave solutions.}
\end{figure}

  Figure \ref{fig:comp0line} shows results for computations of linear
($\lambda=0$) and nonlinear ($\lambda=-2$ and -10) models for $\Psi$
along the $\theta=\pi/2$, $\varphi=0$ lines. All models used rotation
rate $a\Omega=0.3$ and nonlinearity parameter $\Psi_0=0.15$ and the
$180\times20\times32$ grid with an outer boundary at $r=30a$. The
nonlinear models were solved through Newton-Raphson iteration with
continuation in $\lambda$.  The figure shows, as continuous lines, the
computed solutions for outgoing waves with
$\lambda=0,-2,\,-10$. Included in the figure also are the
$\lambda=-2,\,-10$ results for the approximate outgoing waves
extracted from the nonlinear standing wave solutions by the method of
Eq.~(\ref{extracteq}). The difference between these outgoing
approximations and the true outgoing waves is so small that the
approximation results are given as discrete data points to aid in
visualization.

\begin{figure}[ht]
\begin{center}
\includegraphics[width=0.4\textwidth]{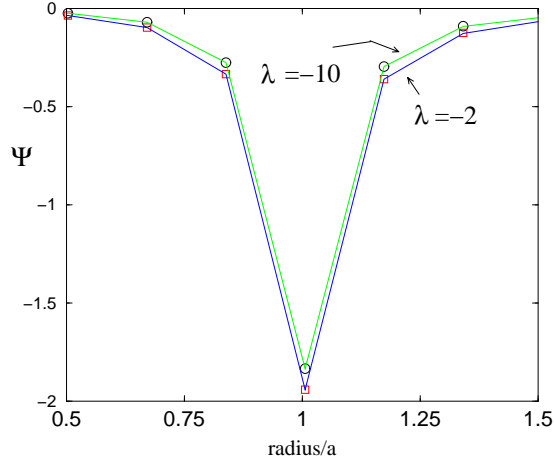}
\end{center}
\caption{\label{fig:compnearsource} 
The same models as in Fig.~\ref{fig:comp0line}, but 
in the region of the sources. As in Fig.~\ref{fig:comp0line}
continuous curves show the computations of the true nonlinear
outgoing waves, and discrete points show the outgoing 
wave approximation extracted from the nonlinear standing wave solution. 
}
\end{figure}

Figure \ref{fig:compnearsource} shows the small-radius portions of the
same models as those in Fig.~\ref{fig:comp0line}.  (The $\lambda=0$
curve is nearly indistinguishable from that for $\lambda=-2$, and is
omitted from the figure.)  The radial $\theta=\pi/2$, $\varphi=0$ line
along which the results are presented, goes through the source point
at $r=a$, so Fig.~\ref{fig:compnearsource} shows the computed solution
in the neighborhood of the source.  The Yukawa-like effect of the
nonlinearity near the source is evident in more rapid fall-off of the
$\lambda=-10$ model away from the source point.

The results in Figs.~\ref{fig:compnearsource} and \ref{fig:comp0line}
are graphical evidence for the accuracy of the PSW method; the
outgoing waveforms extracted from the nonlinear standing wave solution
are excellent approximations to the true outgoing waves both near the
sources and in the wave zone. The agreement in the intermediate zone
(not shown in the figures) is equally impressive.  A quantitative
measure of the agreement is the ``L2'' difference of the outgoing wave
and the extracted outgoing approximation. This measure is the square
root of the average (over all grid points) of the squared difference
between the true and the extracted outgoing solutions. For
$\lambda=-10$ this L2 difference is $8.7\times10^{-6}$ and is of the
same order as the error in Table \ref{table2} for the
$180\times20\times32$ grid being used.  Since the numerical
uncertainties are of the same order as the difference between the true
and the extracted outgoing waves we cannot claim to have computed {\em
any} meaningful inaccuracy in the PSW approximation.

This is unfortunate.  In presenting numerical results it would be
useful to demonstrate that the PSW approximation is, after all, an
approximation by showing a model in which the extracted outgoing
solution is significantly different from the true outgoing
solution. Our inability to do this is related to limitations on
numerical solutions. Our arguments for effective linearity show that
the PSW approximation should fail only if the region of significant
nonlinearity overlaps the wave zone.  For this reason we used the
$\chi$-dependent prefactor of Eq.~(\ref{prevmodelF}) in an earlier
version of the present paper to allow us to force the nonlinearity to
be concentrated in the wave zone. Although that technique did allow us
to induce significant errors in the PSW approximation, we have
explained, following Eq.~(\ref{prevmodelF}), why the solutions for the
models with the prefactor have undesirable features.  If no unnatural
$\chi$-dependence is explicitly injected in the source, the way in
which effective nonlinearity can be made to fail is for $a\Omega$ to
approach unity, i.e.\,, for the source points to move very
relativistically (a case in which the PSW would be expected to fail
for binary inspiral). Unfortunately, we have not been able to find
convergent solutions for large $a\Omega$. Presumably this is due to
the fact that large $a\Omega$ means fields with sharp gradients, too
sharp to be handled by our necessarily coarse grids.

\section{Conclusions}\label{sec:con}

We have given here the foundations of the PSW method based on the
extraction of an outgoing solution from a computed standing-wave
solution. We have also given the details of the extraction
calculation.

The results provided for convergent nonlinear models are ``proof'' by
example that there is no fundamental mathematical problem of
well-posedness of the mixed PDE problem, with radiative boundary
conditions on a sphere that is in both the elliptic and hyperbolic
regions of the problem.  We have, furthermore, presented limited
numerical evidence for the validity of the PSW method, i.e.\,, that
the extracted outgoing solution is a good approximation to the true
nonlinear outgoing solution.  This evidence helps make the case for
the application of the method to the general relativistic problem, in
which only the standing-wave solution will be computable, and the
extracted solution will be taken as the approximation to the physical
problem.

The numerical studies have also taught a lesson about the limitations
of the relatively straightforward numerical method used here, matrix
inversion of the finite difference equations in spherical
coordinates. We have found that this method is limited by the coarse
grid that can be used for the finite-differencing.  We could, in
principle, use a software engineering approach to increase the range
of nonlinearity and rotation rate that can be handled. But the methods
used here, spherical coordinates and delta function sources, are meant
only to provide a relatively simple context for establishing the
foundations for more advanced approaches.

 In a paper now in preparation\cite{paperII}, we will present an
important step forward in dealing with PSW problems, a coordinate
system that conforms to the geometry near the sources and far from the
source asymptotically goes to spherical polar coordinates, the
coordinates best suited to the description of the waves.  One
advantage of this method is that it allows us very simply to put in
details of the sources as inner boundary conditions rather than point
sources. In addition, the new coordinates turn out to be very well
suited to a spectral method that has shown remarkable computational
efficiency, but that poses new computational problems. Computations
using an adapted coordinate system have already been carried out for
the three-dimensional nonlinear scalar problem with both the finite
difference and spectral formulation, and for linearized general
relativity using the finite difference formulation. Since the details
of adapted coordinates, especially with the unusual spectral method,
are not directly related to the foundations of the PSW method, those
details are appropriate to a separate paper.

A very different approach to better numerics is to use  relaxation
methods, already mentioned in Sec.~\ref{sec:num}. In view of the large
number of uncertainties about their application, we have started on a
basic study of relaxation methods in mixed PDE systems in PSW-type
problems, but will continue to explore a number of different
numerical approaches to the PSW problem.

\section{Acknowledgment} We gratefully acknowledge the support of the
National Science Foundation under grants PHY9734871 and PHY0244605. We
also thank the University of Utah Research Foundation for support
during this work.  We thank Christopher Johnson and the
Scientific Computing and Imaging Institute of the University of Utah
for time on their supercomputers to produce the non-FFT results of
Sec. III.  We have also made use of supercomputing facilities
provided by funding from JPL Institutional Computing and Information
Services and the NASA Offices of Earth Science, Aeronautics, and Space
Science. 
We thank John Friedman and Kip Thorne for helpful discussions and
suggestions about this work.
%



\begin{thebibliography}{10}
\bibitem{ctorre} For $\lambda=0$ the problem in Eq.~(\ref{helical3Dscalar})
has been shown to be well posed for a range of source functions and 
boundary conditions; C.~Torre, math-ph/0309008.
\bibitem{prev}
This is an expanded version of an argument given in the two-dimensional context
  in Ref.~\cite{WKP}.
\bibitem{earlierF} This form was used in early versions of the present
paper, is described in Ref.~\cite{safari}, and is closely related to
the nonlinearity used in Ref.~\cite{WKP}.
\end{thebibliography}
\end{document}